\documentclass[letterpaper, 10 pt, conference]{ieeeconf}  

\IEEEoverridecommandlockouts                              

\usepackage[english]{babel}
\usepackage[utf8]{inputenc}
\usepackage[cmex10]{amsmath}
\usepackage{amssymb}
\usepackage{xcolor}
\usepackage{color}

\usepackage[pdftex]{graphicx}
\usepackage{url}
\usepackage{array}

\newcommand{\qed}{\hfill $\blacksquare$}
{}
\newtheorem{corollary}{Corollary}{}
{}
\newtheorem{theorem}{Theorem}{}
\newtheorem{remark}{Remark}{}
\newtheorem{lemma}{Lemma}{}

\title{\LARGE \bf
Constraining Attacker Capabilities Through Actuator Saturation
}

\author{Sahand Hadizadeh Kafash$^{1}$, Jairo Giraldo$^{2}$, Carlos Murguia$^{3}$, Alvaro A.  Cardenas$^{2}$, and Justin Ruths$^{1}$
\thanks{*This work was partially supported by the National Research Foundation (NRF), Prime Minister's Office, Singapore, under its National Cybersecurity R\&D Programme (Award No. NRF2014NCR-NCR001-40) and administered by the National Cybersecurity R\&D Directorate.  Thiby the Air Force Office of Scientific Research under award number FA9550-17-1-0135 and by NSF under award number CNS 1553683.}
\thanks{$^{1}$These authors are with the Departments of Mechanical and Systems Engineering at the University of Texas at Dallas, Richardson, Texas, USA 
        {\tt\small Sahand.HadizadehKafash, jruths @utdallas.edu}}%
\thanks{$^{2}$These authors are with the Computer Science Department at the University of Texas at Dallas, Richardson, Texas, USA
        {\tt\small jag140730, Alvaro.Cardenas @utdallas.edu}}%
\thanks{$^{3}$This author is with the iTrust Center at Singapore University of Technology and Design, Singapore
        {\tt\small murguia\_rendon@sutd.edu.sg}}%
}

\begin{document}

\maketitle
\thispagestyle{empty}
\pagestyle{empty}

\begin{abstract}
For LTI control systems, we provide mathematical tools -- in terms of Linear Matrix Inequalities -- for computing outer ellipsoidal bounds on the reachable sets that attacks can induce in the system when they are subject to the physical limits of the actuators. 
Next, for a given set of dangerous states, states that (if reached) compromise the integrity or safe operation of the system, we provide tools for designing new artificial limits on the actuators (smaller than their physical bounds) such that the new ellipsoidal bounds (and thus the new reachable sets) are as large as possible (in terms of volume) while guaranteeing that the dangerous states are not reachable. This guarantees that the new bounds cut as little as possible from the original reachable set to minimize the loss of system performance. Computer simulations using a platoon of vehicles are presented to illustrate the performance of our tools.
\end{abstract}

\section{INTRODUCTION}

Security and privacy in cyber-physical systems (CPS) have become a major concern in the control community due to tight interaction between communication networks and physical processes \cite{giraldo2017survey}\cite{Cardenas_Sipolini}. 
Several high-profile incidents such as StuxNet \cite{langner2011stuxnet}, the 2016 breach of Ukrainian power grid \cite{zetter2016inside}, as well as attacks on modern cars \cite{checkoway2011comprehensive}   have exposed a wide range of vulnerabilities in CPS.  As a consequence, the use of control techniques to  analyze the impact of cyber-attacks and to design anomaly  detection and mitigation tools have drawn significant attention in recent years \cite{Pasqualetti_1}\nocite{Mo_1}\nocite{Kwon}\nocite{Pappas}\nocite{Gupta}\nocite{Gupta2}\nocite{Cardenas_Sipolini}\nocite{Ahmed1}\nocite{Carlos_Justin2}\nocite{Eric1}-\cite{Cardenas}. 

Most of the work on security analysis does not take into account the physical constraints of actuators; however, it is well known  that  constrained control actions have significant implications in  stability and reachability of control systems \cite{tarbouriech2011stability}.  Since actuators cannot inject arbitrarily large amounts of energy into the system, there are always physical limitations restricting the trajectories that actuators can induce. In most physical dynamical systems actuator saturation arises from these physical limits (e.g., the power that can be injected to an electrical system; the acceleration possible by an engine due to limited torque; the maximum flow rate of an inlet pipe). From a control designer perspective, this translates into a reachability problem: whether it is possible to drive the system from a initial state to a final state given the {actuator bounds}. 

On the other hand, from an attacker's viewpoint and  since attacks on sensors or in control commands lead to anomalous actuators actions, actuator bounds will tend to reduce the adversary capabilities in terms of the states that can be reached by the  the attacker's action. So the question arises: given the actuator bounds, is it possible for the attacker to drive the system to an {undesired} or {dangerous state}? That is, given a set of unsafe states (i.e., the states where the integrity or safe operation of the system is compromised) $\mathcal{D}$, is there a sequence of attacker actions that is capable of driving the system state to $\mathcal{D}$ given the physical restrictions on the actuators? This question reduces to identifying the intersection between the attack-induced reachable set of states and the dangerous state set $\mathcal{D}$. 

The contributions of this work are twofold. First, we approximate the reachable set induced by {individually-bounded control inputs}. Because quantifying this exact set is mathematically intractable, we turn to construct {outer ellipsoidal bounds} of the reachable sets. We provide tools in terms of Linear Matrix Inequalities (LMIs) to obtain tight ellipsoidal bounds for the inherent actuator bounds, using an approach similar to that in \cite{Carlos_Justin3}. Second, we then formulate a design strategy that determines {artificial limits on actuators} smaller than their physical bounds to reduce the potential impact of attackers on the behavior of the system.
To avoid the trivial (and useless) solution of setting the artificial bounds to zero and to minimize the loss of system performance, we find new input bounds that make the new reachable set as large as possible without overlapping a given set of dangerous states $\mathcal{D}$. Effectively our goal is to maximize, through the choice of new actuator bounds, the size of the reachable set without intersecting with unsafe states. In lieu of maximizing the actual reachable set, which is generally intractable, we maximize the volume of the ellipsoidal bounds, phasing this as a synthesis LMI. Intuitively, the ellipsoidal bound is maximized until it touches but does not cross the boundary of the dangerous states.
Finally, we show the viability of our approach in a vehicle  platooning example subject to false-data injection attacks. 


\section{BACKGROUND}

We study Linear Time-Invariant (LTI) discrete-time systems with individually-bounded control inputs:
\begin{equation} \label{eq:LTI}
	x_{k+1} = Fx_k + Gu_k,
\end{equation}
with $k \in \mathbb{N}$; state $x_k \in \mathbb{R}^n$; state matrix $F\in\mathbb{R}^{n\times n}$; input matrix $G\in\mathbb{R}^{n\times m}$; 
and symmetrically bounded control input $u_k \in \mathbb{R}^m$ such that
\begin{equation} \label{eq:actuatorbounds}
	[u_k]_i^2 \leq \gamma_i,\qquad i=1,\dots,m,
\end{equation}
where $\gamma_i>0$ is a constant which determines the bound for the magnitude of each control input, i.e., $[\xi]_j$ is the $j^\text{th}$ element of $\xi$. 

Physical laws and energy constraints lead real control systems to have practical limits on the actuators used to steer the system dynamics. For example, a vehicle cannot accelerate or decelerate infinitely fast; the engine and brakes have limits. These limits imply a saturation in the mapping from input signal to actuation (e.g., once the engine limit is reached, increasing the throttle will not add more torque). These are the input signal limitations modeled by $\gamma_i$.

Unmodeled disturbances propagate through a control system until it comes to effect the input that drives the actuators. Such disturbances include noise and unmodeled forces/dynamics, but also include potential attacks on the control system. \textit{Actuator attacks}\footnote{These attacks can also be accomplished through the installation of malware on the controller hardware.} enter on the communication from controller to actuator, effectively replacing the true controller command with a different one. \textit{Sensor attacks}$^1$ similarly manipulate the measurement information passed from the sensors to the controller. If feedback is employed (which is most often the case), then this corrupted sensor measurements will lead to corrupted input signals to the actuator. 

In this work, we study a generic model that captures any attack that, directly or indirectly, propagates itself to the input signal regardless of the specific mechanism (e.g., feedback law, etc). In particular, we characterize the set of possible reachable states induced by individually-bounded control inputs. We then consider imposing artificial bounds to constrain this reachable set to avoid states that are harmful to the system or unsafe. Because quantifying the exact reachable set is not tractable, we turn to construct outer ellipsoidal bounds on these reachable sets.

In a similar context, the authors in \cite{Reachable_set_1} developed tools to quantify outer ellipsoidal bounds on the reachable set of states for LTI systems with peak bounded input, where the norm of input vector is bounded in aggregate, i.e., $\|u_k\|<\gamma$ (in this paper $\|\cdot\|$ denotes Euclidian norm). We restate their result before we build on it for this work.
\vspace{1mm}
\begin{lemma} \label{lem:that} \cite{Reachable_set_1}.
Let $V_k$ be a positive definite function, $V_1 = 0$, and $\zeta_k^T \zeta_k \leq \kappa \in \mathbb{R}_{>0}$. If there exists a constant $a \in (0,1)$ such that
\begin{equation}\label{Lemma1_Rechable_set_1}
V_{k+1} - aV_k - \frac{1-a}{\kappa} \zeta_k^T \zeta_k \leq 0,
\end{equation}
then, $V_k \leq 1$.
\end{lemma}


\section{RESULTS}
From a theoretical viewpoint, if $(F,G)$ is controllable, the reachable set of \eqref{eq:LTI} is the complete state space $\mathbb{R}^n$ -- even if arbitrarily large inputs are required to reach some states. However, actuators have practical limitations and as a result the entire state-space is not reachable.

Our goal is to find an ellipsoids which encapsulates the entire reachable set. We define these ellipsoids as:
\begin{equation}
\mathcal{E}(P,\alpha):=\left\{ x \in \mathbb{R}^n\ |\ x^T Px\leq \alpha \right\},
\end{equation}
where $P \in \mathbb{R}^{n \times n}$ is a positive-definite matrix and $\alpha \in \mathbb{R}_{>0}$ is a positive constant. When $\alpha=1$, we omit writing it out explicitly, i.e., $\mathcal{E}(P,1) = \mathcal{E}(P)$. We adapt the result in \cite{Reachable_set_1} to help find these ellipsoids.

\vspace{1mm}
\begin{lemma} \label{lem:thatextension}
Let $V_k$ be a positive definite function, $V_1 = 0$, and $[u_k]_i^2 \leq \gamma_i$, $\gamma_i>0$, $i=1,\dots,m$. If there exists a constant $a \in (0,1)$ such that
\begin{equation}\label{eq:lemma}
V_{k+1}-aV_k-(1-a) \sum_{i=1}^{m} \frac{[u_k]_i^2}{\gamma_i} \leq 0,
\end{equation}
then, $V_k \leq m$.
\end{lemma}
\vspace{1ex}
\textbf{\textit{Proof:}} We first simplify \eqref{eq:lemma} bounding it using \eqref{eq:actuatorbounds}
\begin{equation}
	V_{k+1} \leq aV_k + (1-a) \underbrace{\sum_{i=1}^{m} \frac{[u_k]_i^2}{\gamma_i}}_{\leq m}.
\end{equation}
Because of inequality \eqref{eq:lemma}, the following is satisfied
\begin{equation}\label{eq:lemmab}
	V_k \leq aV_{k-1} + (1-a)m,
\end{equation}
substituting \eqref{eq:lemmab} in \eqref{eq:lemma} and continuing the recursion yields
\begin{align*}
V_{k} &\leq aV_{k-1} + (1-a)m \\
&\leq a\left[ aV_{k-2} + (1-a)m \right] + (1-a)m \\
&= a^2V_{k-2} + (1-a^2)m \\
&\quad\vdots \\
&\leq a^{k-1}V_1 + (1-a^{k-1})m.
\end{align*}
Since $a \in (0,1) 1$ and $V_1=0$, $V_k \leq m$ for all $k \geq 1$. \qed

\subsection{Analysis}
We can now employ Lemma \ref{lem:thatextension} to derive an outer ellipsoidal bound on the reachable set of \eqref{eq:LTI} subject to {known individual actuator bounds} \eqref{eq:actuatorbounds}. The reachable set we seek to quantify is given by
\begin{equation}\label{Reach1}
	\mathcal{R} := \left\{ x_k\in\mathbb{R}^n\ \Bigg|\ 
    \begin{aligned}
		&x_{k+1} = Fx_k + Gu_k,\ x_1 = \mathbf{0}, \\
        &[u_k]_i^2 \leq \gamma_i,\hspace{.5mm}  i=1,\ldots,m,\hspace{.5mm} \forall\hspace{.5mm} k\in\mathbb{N}
	\end{aligned}  \right\}.
\end{equation}
Notice that
\begin{equation}
	\sum_{i=1}^{m} \frac{[u_k]_i^2}{\gamma_i} = u_k^TRu_k \leq m,
\end{equation}
where the actuator bounds are collected in the matrix R,
\begin{equation} \label{eq:R}
R:=
\begin{bmatrix}
\frac{1}{\gamma_1} & 0 & 0
\\
0 & \ddots & 0
\\
0 & 0 & \frac{1}{\gamma_m}
\end{bmatrix}.
\end{equation}

\begin{remark}
Note, from \eqref{Reach1}, that if for some $k=k^*$, $x_{k^*} \neq 0$ and $\rho[F] > 1$, where $\rho[\cdot]$ denotes spectral radius, then $||x_{k}||$ diverges to infinity as $k$ grows for any non-stabilizing $u_k$. That is, $\mathcal{R}$ is \textit{unbounded} if the system is \textit{open-loop unstable}. If $\rho[F] \leq 1$, then $||x_{k}||$ may or may not diverge to infinity depending on algebraic and geometric multiplicities of the eigenvalues with unit modulus of $F$ (a known fact from stability of LTI systems), see \cite{Astrom} for details.
\end{remark}

\vspace{1mm}
\begin{theorem} \label{thm:analysis}
Consider the LTI system \eqref{eq:LTI} with matrices $(F,G)$, the actuator bounds $\gamma_i>0$, $i=1,\dots,m$, and matrix $R$ in \eqref{eq:R}. For given  $a\in(0,1)$, if there exists a positive definite matrix $P \in \mathbb{R}^{n \times n}$ solution of the following convex optimization:
\begin{equation} \label{eq:analysis}
\left\{\begin{aligned}
	&\min_{P}\ -\log\det{P},\\
    &\text{s.t.}\ P>0,\ \text{and}\\
    &\quad \begin{bmatrix}
		aP - F^T PF & -F^T PG \\ -G^T PF & (1-a)R - G^T PG
	 \end{bmatrix} \geq 0,
\end{aligned}\right.
\end{equation}
then, $\mathcal{R} \subseteq \mathcal{E}(P,m)$ and the ellipsoid $\mathcal{E}(P,m)$ has minimum volume.
\end{theorem}
\vspace{1ex}
\textbf{\textit{Proof:}} For some positive definite matrix $P\in\mathbb{R}^{n\times n}$, let $V_k=x_k^TPx_k$ in Lemma \ref{lem:thatextension}. Substituting \eqref{eq:LTI} and this $V_k$ in \eqref{eq:lemma}  yields
\begin{equation} \label{eq:analysis_LMI}
\nu^T
\underbrace{\begin{bmatrix}
aP - F^T PF & -F^T PG \\ -G^T PF & (1-a)R - G^T PG 
\end{bmatrix}}_{Q}
\nu \geq 0
\end{equation}
where $\nu=\left[x_k^T,\ u_k^T\right]^T$. This inequality is satisfied if and only if $Q$ is positive semi-definite.

To ensure that the ellipsoid bound is as tight as possible, we minimize $(\det{P})^{-1/2}$ since this quantity is proportional to the volume of $x_k^TPx_k=m$. We instead minimize $\log\det{P^{-1}}$ as it shares the same minimizer and because for $P>0$ this objective is convex \cite{BEFB:94}. \qed

\vspace{1mm}
Lemma \ref{lem:thatextension} indicates that the solution to the optimization problem \eqref{eq:analysis} may exist for some values of the parameter $a\in(0,1)$. Out of these values, we are interested in selecting the one that leads to the ellipsoid with minimum volume. We employ a straightforward grid search to find this value of $a$.
 
\vspace{1mm}
\begin{remark} \label{rmk:samebounds}
In the case in which all control inputs have identical bounds, i.e., $R = \frac{1}{\gamma} I_{m}$, with $I_m$ the $m \times m$ identity matrix, the common scalar bound $\gamma$ can be factored out of the LMI in \eqref{eq:analysis_LMI} by defining $\hat{P}=\gamma P$, then:
\begin{equation} \label{eq:samebounds}
	Q = \frac{1}{\gamma}
	\underbrace{\begin{bmatrix}
		a\hat{P} - F^T \hat{P}F & -F^T \hat{P}G \\ -G^T \hat{P}F & (1-a)I_m - G^T \hat{P}G 
	\end{bmatrix}}_{\hat{Q}} \geq 0.
\end{equation}
Since $\gamma>0$, $\hat{Q}\geq 0 \Leftrightarrow Q \geq 0$. This implies that in the case of common actuator bounds, the optimization can be solved independent of the actual bound. After a solution $\hat{P}$ satisfying $\hat{Q}\geq 0$ in \eqref{eq:samebounds} is found, it is simply scaled by the bound $\gamma$ to recover the desired ellipse, i.e., $\mathcal{R} \subseteq \mathcal{E}\left(\frac{1}{\gamma} \hat{P},m\right)$.
\end{remark}

\subsection{Synthesis}
In most physical dynamical systems, actuator saturation arises from physical limits. In the past section, Theorem \ref{thm:analysis} gives us the tools necessary to quantify the outer ellipsoidal bounds for the reachable states according to such inherent actuator bounds. In the context of security, it is intriguing to impose artificial limits on actuators smaller than their physical bounds to reduce the potential impact of attackers on the behavior of the system. Such a design problem would be informed by a region of state space $\mathcal{D}$ which is considered unsafe. Such a region might represent states in which, for example, the pressure of a holding vessel will exceed its pressure rating or the level of a liquid in a tank exceeds its capacity. Our aim is that, through the selection of new input bounds, we can guarantee that the system would avoid these \textit{dangerous states}, not simply due to stabilizing controller action, which might be hacked, but due to the imposed new limits of the actuator action.

Thus, we aim here to design new bounds $\hat{\gamma}_i$, $i=1,\ldots,m$ such that the new reachable set bounding ellipsoid does not overlap with the unsafe states. Corresponding to these new bounds, we define the new rechable set $\hat{\mathcal{R}}$ as:
\begin{equation}\label{Rhat}
	\hat{\mathcal{R}} := \left\{ x_k\in\mathbb{R}^n\ \Bigg|\ 
    \begin{aligned}
		&x_{k+1} = Fx_k + Gu_k,\ x_1 = \mathbf{0}, \\
        &[u_k]_i^2 \leq \hat{\gamma}_i,\hspace{.5mm}  i=1,\ldots,m,\hspace{.5mm} \forall\hspace{.5mm} k\in\mathbb{N}
	\end{aligned}  \right\}.
\end{equation}

The dangerous state sets in many, if not most, practical applications can be captured through the union of half-spaces defined by their boundary hyperplanes:
\begin{equation}
	\mathcal{D} := \left\{ x\in\mathbb{R}^n\ \Bigg|\ \bigcup_{i=1}^\kappa c_i^Tx\geq b_i  \right\},
\end{equation}
where each pair $(c_i,b_i)$, $c_i\in\mathbb{R}^n$, $b_i\in\mathbb{R}$, $i=1,\dots,\kappa$ quantifies a hyperplane that defines a single half-space.

\vspace{1mm}
\begin{theorem} \label{thm:synthesis} Consider the LTI system \eqref{eq:LTI} with matrices $(F,G)$, {the original actuator bounds} $\gamma_i>0$, $i=1,\dots,m$, matrix $R$ in \eqref{eq:R}, and a set $\mathcal{D}$ of dangerous states bounded by the hyperplanes $c_i^T x = b_i$, $i=1,\dots,\kappa$. For given $a\in(0,1)$,\linebreak if there exist a positive definite matrix $Y \in \mathbb{R}^{n \times n}$ and diago\-nal matrix $\hat{R}:= \text{diag}(\hat{r}_1,\ldots,\hat{r}_m) \in \mathbb{R}^{m \times m}$, $\hat{r}_i>0$, solution of the following convex optimization:
\begin{equation} \label{eq:synthesis}
	\left\{\begin{aligned}
    &\min_{\hat{R},Y}\ \text{tr}(\hat{R}), \\
    &\text{s.t. } \hat{R} \geq R,\ Y > 0, \\
    &\quad\ c_i^T Y c_i \leq \frac{b_i^2}{m},\quad \text{for  }i=1\ldots,\kappa, \\
    &\quad \begin{bmatrix}
    a Y & 0 & YF^{T}\\
    0 & (1-a)\hat{R} & G^{T}\\
    F Y & G & Y
    \end{bmatrix} \geq 0,
	\end{aligned}\right.
\end{equation}
then, the {new actuator bounds} $\hat{\gamma_i}:=(1/\hat{r}_i)$, $i=1,\dots,m$, enforce that $\mathcal{D}$ does not intersect with the new reachable set $\hat{\mathcal{R}}$ in \eqref{Rhat} and {maximize the volume} of the {new minimum-volume ellipsoid} $\mathcal{E}(Y^{-1},m)$ bounding $\hat{\mathcal{R}}$. 
\end{theorem}
\vspace{1ex}
\textbf{\textit{Proof:}} The minimum distance $d$ between an ellipsoid $\mathcal{E}(P,m)$ centered at zero and a hyperplane $c^Tx=b$ is given by the formula \cite{Kurzhanski}
\begin{equation}
	d = \frac{|b| - \sqrt{m c^T P^{-1} c}}{\sqrt{c^T c}}.
\end{equation}
We aim to obtain the largest ellipsoid (in terms of volume) that does not cross the hyperplane. This would maximize the size of the reachable set restricted to not crossing into the dangerous set. This is accomplished if we let the ellipsoid and hyperplane to touch at a single point, i.e., for the distance to be zero, which implies
\begin{equation} \label{eq:distance_constraint_single}
	c^TP^{-1}c = \frac{b^2}{m}.
\end{equation}
Given the hyperplane parameters $c$ and $b$, the choice of $P^{-1}$ that satisfies this relationship is then the largest ellipse $\mathcal{E}(P,m)$ that does not overlap with the dangerous states. Note that the zero-distance condition \eqref{eq:distance_constraint_single} is written in terms of $P^{-1}$ which is not linear in $P$. Thus, to maintain a tractable convex semi-definite optimization problem, we write the original analysis LMI $Q$ in \eqref{eq:analysis_LMI} in terms of $Y:=P^{-1}$ and the new matrix $\hat{R}$. The new $Q$ can be written as the Schur complement of a higher dimensional matrix $\widetilde{Q}$ such that the positive semi-definiteness of $\widetilde{Q}$ implies the positive semi-definiteness of $Q$,
\begin{equation}
Q \geq 0 \Leftrightarrow \widetilde{Q}=
\begin{bmatrix}
aY^{-1} & 0 & F^T Y^{-1}\\
0 & (1-a) \hat{R} & G^T Y^{-1}\\
Y^{-1}F & Y^{-1}G & Y^{-1}
\end{bmatrix}
\geq 0.
\end{equation}
Finally, multiplying $\widetilde{Q}$ above from the left and right by the following congruence transformation
\begin{equation}\label{congruence}
\mathcal{Y} :=	\begin{bmatrix} Y&&\\&I_{n}&\\&&Y \end{bmatrix},
\end{equation}
results in the bottom LMI in \eqref{eq:synthesis}. 

The constraint \eqref{eq:distance_constraint_single} can be added for each hyperplane that specifies the boundary of the dangerous set. Once multiple hyperplanes specify the boundary, it is possible that the ellipse cannot touch all boundaries simultaneously; therefore, we relax each distance constraint into the inequality in \eqref{eq:synthesis} ($c^TP^{-1}c = c^TYc \leq \frac{b^2}{m}$). Taken together, these distance constraints ensure that the outer ellipsoidal bound of reachable states does not extend beyond any of the hyperplanes. In order to guarantee that we find the largest possible bounds, we maximize a function of the bounds, or in this case minimize a function of the diagonal elements of $\hat{R}$ (the trace of $\hat{R}$). This bound maximization coupled with the hyperplane boundaries have the additional effect of ensuring the tightness ellipsoidal bound (minimum volume) to the reachable set. This allows us to omit the $\log\det P^{-1}$ objective we use in Theorem \ref{thm:analysis} (which is not a convex objective with the decision variable as $Y=P^{-1}$). Maximizing the bounds first expands the ellipse until it hits one or more hyperplanes. Then, continuing to maximize the bounds maintains the same ellipsoid, however, increases the tightness of the ellipsoid onto the reachable states, until this cannot be improved. The restriction $\hat{R}\geq R$ is added to ensure that the new bounds are not larger than the original physical bounds. \qed

\vspace{1mm}
A particular case in which a close-form expression for the new actuator bounds can be obtained is when all the bounds are designed to be equal, i.e., $\hat{\gamma_i}:=\hat{\gamma} \in \mathbb{R}_{>0}$, $i=1,\dots,m$. This is stated in the following corollary of Theorem 2.

\vspace{1mm}
\begin{corollary}\label{cor:samebound}
Let $\gamma_{\min} := \min(\gamma_i)$, $i=1,\dots,m$. For given $a\in(0,1)$, if there exists a positive definite matrix $\hat{P} \in \mathbb{R}^{n \times n}$ solution of the following convex optimization:
\begin{equation} \label{eq:synthesis2}
\left\{\begin{aligned}
	&\min_{\hat{P}}\ -\log\det{\hat{P}},\\
    &\text{s.t.}\ \hat{P}>0,\ \text{and}\\
    &\quad \begin{bmatrix}
      a\hat{P} - F^T \hat{P}F & -F^T \hat{P}G \\ -G^T \hat{P}F & (1-a)I_m - G^T \hat{P}G
\end{bmatrix} \geq 0,
\end{aligned}\right.
\end{equation}
then, the {new actuator bound}:
\begin{equation} \label{eq:synthesis3}
\hat{\gamma}:= \min\left(\frac{b_i^2}{mc_i^T\hat{P}^{-1}c_i}, \gamma_{\min} \right), \hspace{1mm} i=1,\dots,m,
\end{equation}
enforce that $\mathcal{D}$ does not intersect with the new reachable set $\hat{\mathcal{R}}$ in \eqref{Rhat} with $\hat{\gamma_i}=\hat{\gamma}$ and {maximize the volume} of the {new minimum-volume ellipsoid} $\mathcal{E}(\frac{1}{\hat{\gamma}}\hat{P},m)$ bounding $\hat{\mathcal{R}}$.
\end{corollary}
\vspace{1ex}
\textbf{\textit{Proof:} }
Because $\hat{\gamma_i}=\hat{\gamma}$, we have $\hat{R} = \frac{1}{\hat{\gamma}}I_m$ in \eqref{eq:synthesis}. Applying the congruence transformation $\mathcal{Y}^{-1}$ (with $\mathcal{Y}$ in \eqref{congruence}), the Schur complement, and the change of variables $P=Y^{-1}$ to the bottom LMI in \eqref{eq:synthesis}, we get the equivalent inequality
\begin{equation*}
\frac{1}{\hat{\gamma}}\begin{bmatrix}
      a\hat{P} - F^T \hat{P}F & -F^T \hat{P}G \\ -G^T \hat{P}F & (1-a)I_m - G^T \hat{P}G
\end{bmatrix} \geq 0,
\end{equation*}
with $\hat{P}=\hat{\gamma} P$. Which is the bottom inequality in \eqref{eq:synthesis2} scaled by $\frac{1}{\hat{\gamma}}$. This implies that for a solution $\hat{P}$ of \eqref{eq:synthesis2}, we can simply scale $\hat{P}$ by the inversed new bound $\hat{\gamma}$ to recover the desired ellipse, i.e., $\hat{\mathcal{R}} \subseteq \mathcal{E}(\frac{1}{\hat{\gamma}} \hat{P},m)$. Note that $\mathcal{E}(\frac{1}{\gamma} \hat{P},m)$ is a minimum-volume ellipsoid for any $\hat{\gamma}>0$ because $\mathcal{E}(\hat{P},m)$ is minimum-volume and both ellipsoids have the same shape and orientation (i.e., the eigenvectors of $P$ and $\hat{P}$ are the same). Hence, to have a maximal minimum-volume ellipsoid, we have to select $\hat{\gamma}$ to be the largest $\hat{\gamma}$ such that the dangerous states are avoided.  For this, as in Theorem 2, we use the distance condition $c_i^T Y c_i = c_i^T P^{-1} c_i = \hat{\gamma} c_i^T \hat{P}^{-1}c_i  \leq \frac{b_i^2}{m}$,\linebreak $i=1,\ldots,m$. The largest $\hat{\gamma}$ satisfying the latter inequality and being smaller than the smallest original physical bound is given by $\hat{\gamma}$ in \eqref{eq:synthesis3}. \qed


\begin{figure*}[t]
\centering
\includegraphics[width=1.0\linewidth]{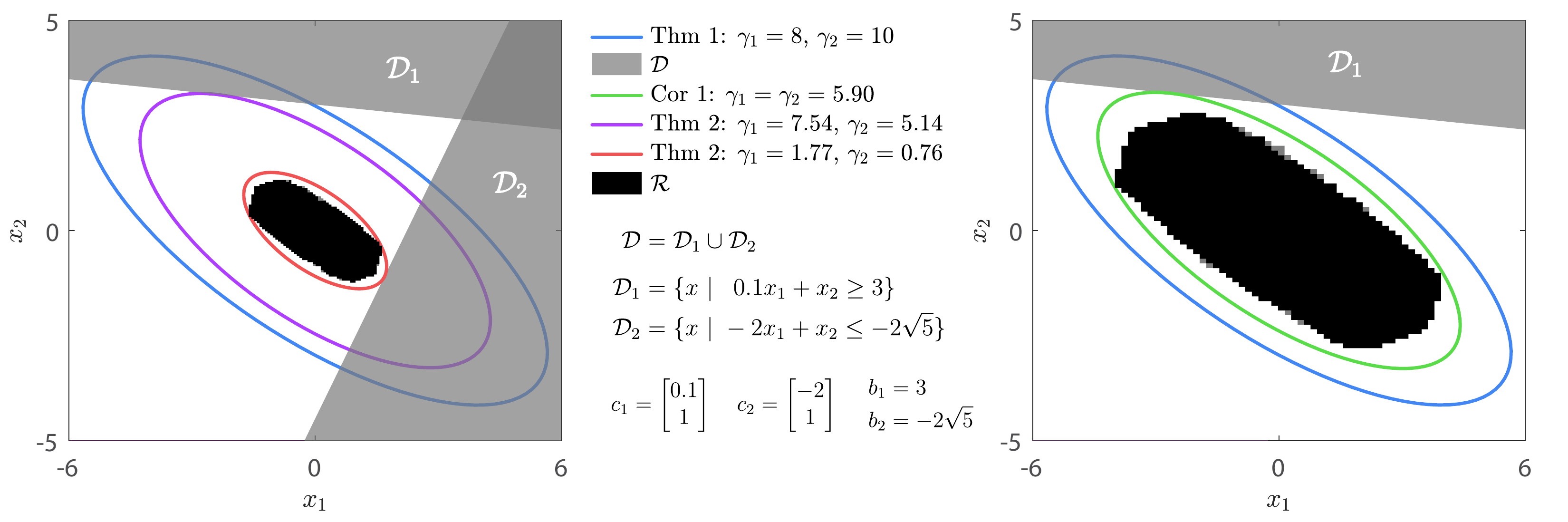}
\caption{The solution of Theorem \ref{thm:analysis} provides the (blue) ellipsoid which outer-bounds the reachable set corresponding to the system driven by the original physical input bounds $\gamma=[8,10]$. A dangerous state set $\mathcal{D}$, is defined by the single half-space $c_1^Tx\geq b_1$ and the solution of Theorem \ref{thm:synthesis} provides the bounds $\gamma=[7.54,5.14]$ and corresponding (purple) ellipsoid that outer-bounds the largest reachable set that avoids the dangerous state set. A second half-space is added to $\mathcal{D}$, $c_2^Tx\leq b_2$ and Theorem \ref{thm:synthesis} again provides the bounds $\gamma=[1.77,0.76]$ and the (red) ellipsoid that outer-bounds the largest reachable set that avoids both half-spaces. Corollary \ref{cor:samebound} can be used to derive equal bounds $\gamma=[5.9,5.9]$ and the corresponding (green) ellipsoid that bound the reachable set that avoids the single half-space $c_1^Tx\geq b_1$. We show a Monte-Carlo simulation of many (10,000) trajectories to construct an empirical reachable set according to the corresponding bounds, which demonstrates the tightness of the ellipsoidal bounds.}\label{fig:illustrative}
\end{figure*}

\section{ILLUSTRATIVE EXAMPLE}
To illustrate these results, we consider a linear $2\times 2$ system with two inputs
\begin{equation}
F=
\begin{bmatrix}
0.84 & 0.23 \\ -0.47 & 0.12
\end{bmatrix},\qquad
G=
\begin{bmatrix}
0.07 & 0.3 \\ 0.23 & 0.1
\end{bmatrix}.
\end{equation}
The actuators of this system have physical limitations which saturate the inputs at $[u_k]_1^2\leq\gamma_1=8$ and $[u_k]_2^2\leq\gamma_2=10$, $k\in\mathbb{N}$. These inherent bounds impose a reachable set of states which are outer-bounded by an ellipse that is the solution of Theorem \ref{thm:analysis}. This ellipse is the blue ellipse in Fig. \ref{fig:illustrative}. This solution (and all others we present here) is found using YALMIP, a MATLAB toolbox for optimization, especially semidefinite programs (SDP), and the SDP solver SeDuMi \cite{YALMIP},\cite{sedumi}. As mentioned before, the parameter $a$ is incrementally varied over the range $(0,1)$ and the optimization is solved at each value of $a$. The optimal solution must first be a successful solution (for some values of $a$ there is no solution to the problem) and second must yield an ellipse of minimum volume. 

We now define a set of dangerous states that represent states which jeopardize the safety and/or operation of the system. Let $\mathcal{D}=\mathcal{D}_1=\{x\ |\ 0.1x_1+x_2\geq 3\}$. It is immediately apparent that the inherent bounds allow the system to reach some of the dangerous states. Using Theorem \ref{thm:synthesis}, we can find new artificial bounds on the inputs so that the system can no longer reach the danger states. We solve the optimization for the purple ellipse in Fig. \ref{fig:illustrative} corresponding to bounds $\gamma_1=7.54$ and $\gamma_2=5.14$. This outer ellipsoidal bound touches, but does not cross the hyperplane that defines $\mathcal{D}_1$. Thus, following these new bounds, the system cannot reach any of the dangerous states.

It is also possible to avoid the dangerous states while enforcing equivalent artificial bounds on all inputs. Using Corollary \ref{cor:samebound}, we identify the ellipsoid $\hat{P}$. Using \eqref{eq:synthesis3}, we find the bound $\gamma_1=\gamma_2=\gamma=5.9$ and the corresponding green ellipsoid  in Fig. \ref{fig:illustrative}. This ellipsoid (as well as the others above) tightly bound the reachable set. We show this by plotting an extensive Monte-Carlo simulation of 10,000 trajectories (each of length 1000 steps) of the system with the bounds $\gamma_1=\gamma_2=\gamma=5.9$. This empirical reachable set is well approximated by the outer bounding ellipsoid.

We further demonstrate that the formulation in Theorem \ref{thm:synthesis} can handle a dangerous state set composed of the union of multiple half-spaces. Let $\mathcal{D}=\mathcal{D}_1\cup\mathcal{D}_2$, where $\mathcal{D}_2 = \{x\ |\ -2x_1+x_2\leq -2\sqrt{5}\}$. Theorem \ref{thm:synthesis} with both half-space constraints produces the bounds $\gamma_1=1.77$ and $\gamma_2=0.76$ along with the corresponding red ellipse in Fig. \ref{fig:illustrative}. Again we demonstrate the tightness of this ellipsoidal bound by plotting the empirical reachable set.

\section{CASE STUDY: PLATOONING}
\begin{figure}[t]
\centering
\includegraphics[width=\columnwidth]{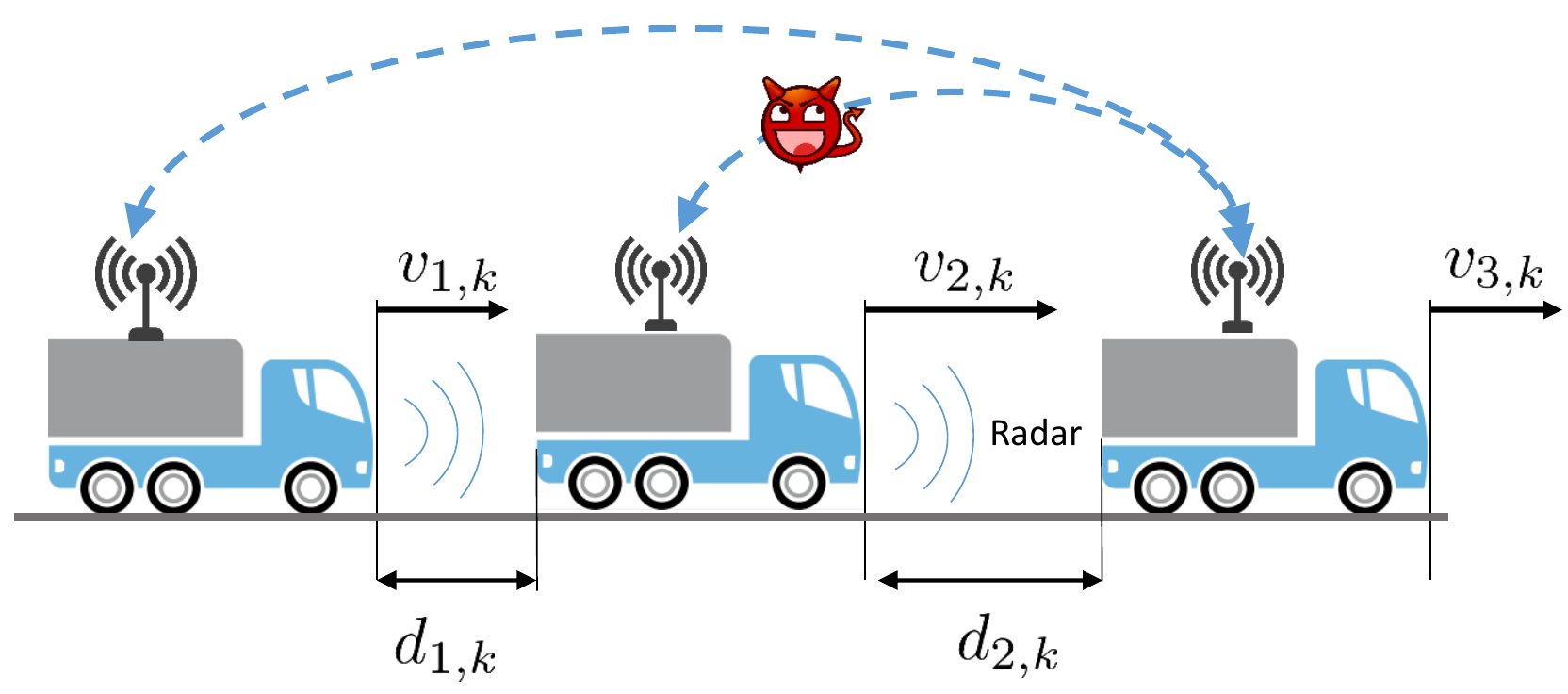}
\caption{Scheme of a platoon of three vehicles. Each vehicle can sense adjacent distances and speeds using radar/lidar. In addition, they are also equipped with a CACC strategy using a, e.g., vehicle-to-vehicle, communication network.  An adversary can gain access to some sensors or actuator commands transmitted through the network.  }\label{fig:platooning_scheme}
\end{figure}

In order to illustrate the viability of our analysis, we  consider the platooning problem depicted in Figure \ref{fig:platooning_scheme}.  In particular, platooning offers many benefits  over solo driving such as better reaction times, decrease of CO$_2$ emissions, and lower fuel consumption \cite{suthaputchakun2012applications}. The objective of the platoon is to maintain an adequate distance between vehicles, such that sudden changes in the leader's speed (e.g., braking) will not cause any crash in the preceding vehicles. This is known as the string stability of the platoon and has been widely studied in the literature \cite{swaroop1996string,ploeg2014controller,oncu2014cooperative}.  Typically, the Adaptive Cruise Control (ACC) system controls the distance and/or relative velocity between adjoining vehicles by measuring (radar/lidar) and reacting to the relative distance and/or velocity between adjacent vehicles compared to a desired setpoint. More recently, work has leveraged vehicle-to-vehicle or infrastructure-to-vehicle communication to inject feed-forward commands. Such Cooperative Adaptive Cruise Control (CACC) systems improve the string stability of the platoon and allows vehicles to follow each other with a closer distance than with ACC, thereby improving traffic flow capacity. CACC gathers information of vehicles further in front according to a specific communication network topology. 

Let us consider a simplified discrete-time cruise control model of a   platoon with $n$ vehicles as follows
\begin{align}
d_{1,k+1}&=d_{1,k}+\Delta_t (v_{2,k}-v_{1,k})\nonumber\\
&\ \vdots\\
d_{n-1,k+1}&=d_{n-1,k}+\Delta_t(v_{n,k}-v_{n-1,k})\nonumber\\
v_{1,k+1}&=v_{1,k}+\beta_1 v_{1,k}+\Delta_tu_{1,k}\nonumber\\
&\ \vdots\\
v_{n,k+1}&=v_{n,k}+\beta_n v_{n,k}+\Delta_tu_{n,k}
\end{align}
where $k\in\mathbb{Z}_+$ is the sampling instant, $d_{i,k}$ is the distance between vehicle $i+1$ and $i$ for  $i=1,\ldots,n-1$, $v_{j,k}$ is the  speed of the $j^{th}$ vehicle, and $u_{j,k}$ is the control input that changes the acceleration, for $j=1,\ldots,n$.  $\Delta_t$ is the sampling period and $\beta_j<0$ is the velocity loss caused by friction. For simplicity, let $d_k=[d_{1,k},\ldots,d_{n-1,k}]$, $v_k=[v_{1,k},\ldots,v_{n,k}]$, and $u_k=[u_{1,k},\ldots, u_{n,k}]$. As depicted in Fig. \ref{fig:platooning_scheme}, the leader vehicle is indexed by $n$ and the last vehicle is index $1$.
 

For our example, we consider a simple  control strategy that combines an ACC and a secondary control (e.g., CACC) given by
\[u_{i,k}=\widetilde u_{i,k}+w_{i,k}(d_k,v_k),\]
 where $\widetilde u_{i,k}$ corresponds to a forward-and-reverse-looking proportional-derivative control  according to \cite{dadras2015vehicular} of the form 
\begin{align}
\widetilde u_{i,k}=k_{p,i}(-d_{i-1,k}+d_{i-1}^*) +k_{p,i}(d_{i,k}-d_i^*)\nonumber\\
+k_{d,i}(v_{i-1,k}-v_{i,k}) +k_{d,i}(v_{i,k}-v_{i+1,k}),\end{align}
 to maintain a desired distance $d_i^*$ between vehicles and $w_{i,k}(d_{k},{v}_k)$ is the secondary control strategy that relies on a  communication network. However, as depicted in Figure \ref{fig:platooning_scheme}, an adversary can intercept and modify that information causing dangerous impacts such as making the vehicles crash. 

Because of the physical constraints in the acceleration/deceleration of each vehicle, we assume that the secondary control action is bounded according to 
\begin{align}
\underline w_{i}\leq w_{i,k}\leq \bar w_{i}. 
\end{align}
for $\underline w_{i}<\bar w_{i}$. As we mentioned above, we can impose virtual constraints in control actions in order to avoid unsafe states. 

\subsection{Experiments}
\begin{figure}[t]
\centering
\includegraphics[width=\columnwidth]{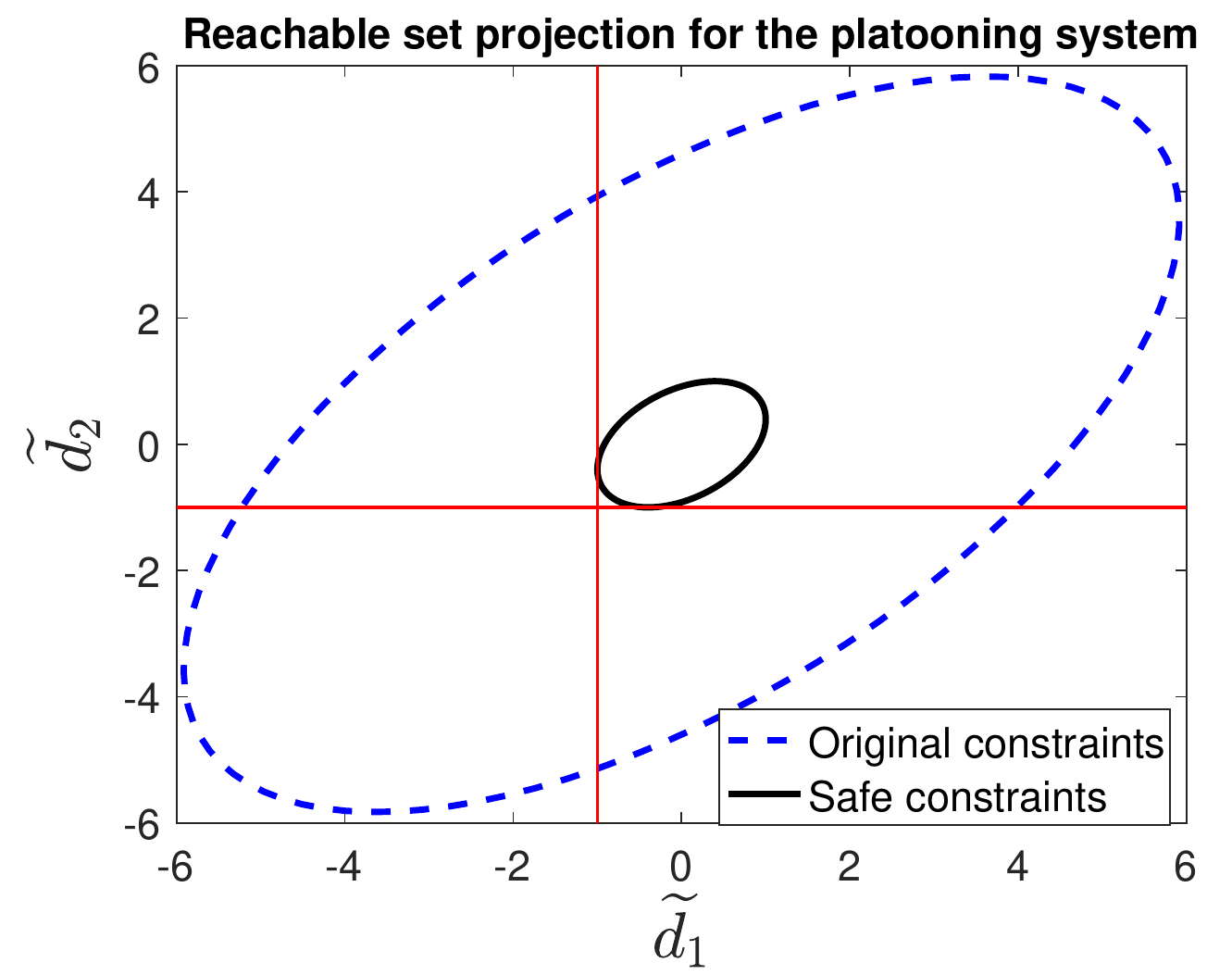}
\caption{Projection of the reachable set in the $\widetilde d_1-\widetilde d_2$ plane for the platooning example with unsafe hyperplanes $\widetilde d_i\leq -1$ for $i={1,2}$. Notice that with  the original constraints the unsafe states can be reached, but   imposing  the safe constraints obtained using our results ensures that no disturbance or attack will drive the system to unsafe states.}\label{fig_experiment1}
\end{figure}

Suppose we have a platoon of three vehicles as depicted in Fig. \ref{fig:platooning_scheme} where  $d_i^*$ and $v^*$ are the desired separation distance and desired velocity, respectively. We can introduce a change of variable  $\widetilde d_{i,k}=d_{i,k}-d_i^*$ and $\widetilde v_{i,k}=v_{i,k}-v^*$  without affecting the dynamic model,  such that $\widetilde d_{i,k}=0$ implies that the desired reference is achieved, i.e.,  $d_i=d_i^*$. Therefore, let ${x}_k=[\widetilde d_{1,k},\widetilde d_{1,k},\widetilde v_{1,k},\widetilde v_{2,k},\widetilde v_{3,k}]^\top$ 
such that 
\[\tiny
F=\begin{bmatrix}
1 & 0 & -\Delta_t & \Delta_t & 0\\
0 & 1 & 0&  -\Delta_t & \Delta_t\\
k_{p,1} & 0 &\!\!\!\! (1+\beta_1)-k_{d,1}\!\!\! & k_{d,1} & 0\\
-k_{p,2} & k_{p,2} & k_{d,2}&\!\!\!  (1+\beta_2)-2k_{d,2}\!\!\! & k_{d,2}\\
0 & -k_{p,3} & 0 & k_{d,3} &\!\!\! (1+\beta_3)-k_{d,3}\\
\end{bmatrix},
\]
\[
G=\begin{bmatrix}\boldsymbol{0}_{2\times 3}\\
\Delta_t I_3
\end{bmatrix}.
\]
The dynamic system with the ACC and the secondary control $w_k$ is of the form 
\[x_{k+1}=Fx_k+Gw_k.\] 
Notice that  $\widetilde d_{i,k}=-d_i^*$ corresponds to the case when   the distance $d_{i,k}=0$,  which means that   the pair of vehicles $i$ and $i+1$ have  crashed. Since we want to avoid crashes, we define  the unsafe states as 
\begin{equation*}
	\mathcal{D} = \left\{ {x}\in\mathbb{R}^n\ \big|\ -x_1\geq d_1^* \cup -x_2\geq d_2^*  \right\}.
\end{equation*}
The simulation parameters are  $d_i^*=d^*=1\;$m, $v^*=60\;$km/h, $\Delta_t=0.5\;$s, $\beta_i=-0.1$,  $k_{p,i}=0.2$, $k_{d,i}=0.3$. The initial acceleration constraints are   $-\underline w_{1}=\bar w_{1}=1.1\; \text{m/s}^\text{2}$, $-\underline w_{2}=\bar w_{2}=\;0.9\; \text{m/s}^\text{2}$, and $-\underline w_{3}=\bar w_{3}=1.05 \;\text{m/s}^\text{2}$, such that $\gamma=(1.2,0.8,1.1)$ (the square of the absolute bounds).  Figure \ref{fig_experiment1} illustrates the  projection in the $\widetilde d_1-\widetilde d_2$ space of the minimum volume ellipsoid that approximates the  reachable set  using Theorem \ref{thm:analysis}  for the given acceleration constraints.

However, notice that unsafe states may be reached and thus there are inputs that can lead the vehicles to crash.   Applying Theorem \ref{thm:synthesis},    we are able to find the set of constraints that will keep the reachable states outside of the dangerous states, as depicted in Fig. \ref{fig_experiment1}. The new set of bounds (safe constraints) is  $\gamma=(0.03,0.05,0.03)$, such that any sensor or actuator attack that affects the control input will not be able to cause the vehicles to crash.\\
Let us consider a simple CACC  strategy of the form 
\[ w_{k}=K{x}_{k}
\]
where $K$ is a LQR control gain. This control strategy  drives ${x}_{k}$ to $0$ as $k\rightarrow \infty$, such that the intervehicle distances and velocities become $d_i=1\;$m and $v_j=60\;$km/h for all $i,j$. The CACC  will gather information about  vehicles position and speed using wireless communications (e.g., vehicle-to-vehicle) with full information, i.e., each vehicle has access to  all the states.  

\begin{figure}[t]
\centering
\includegraphics[width=\columnwidth]{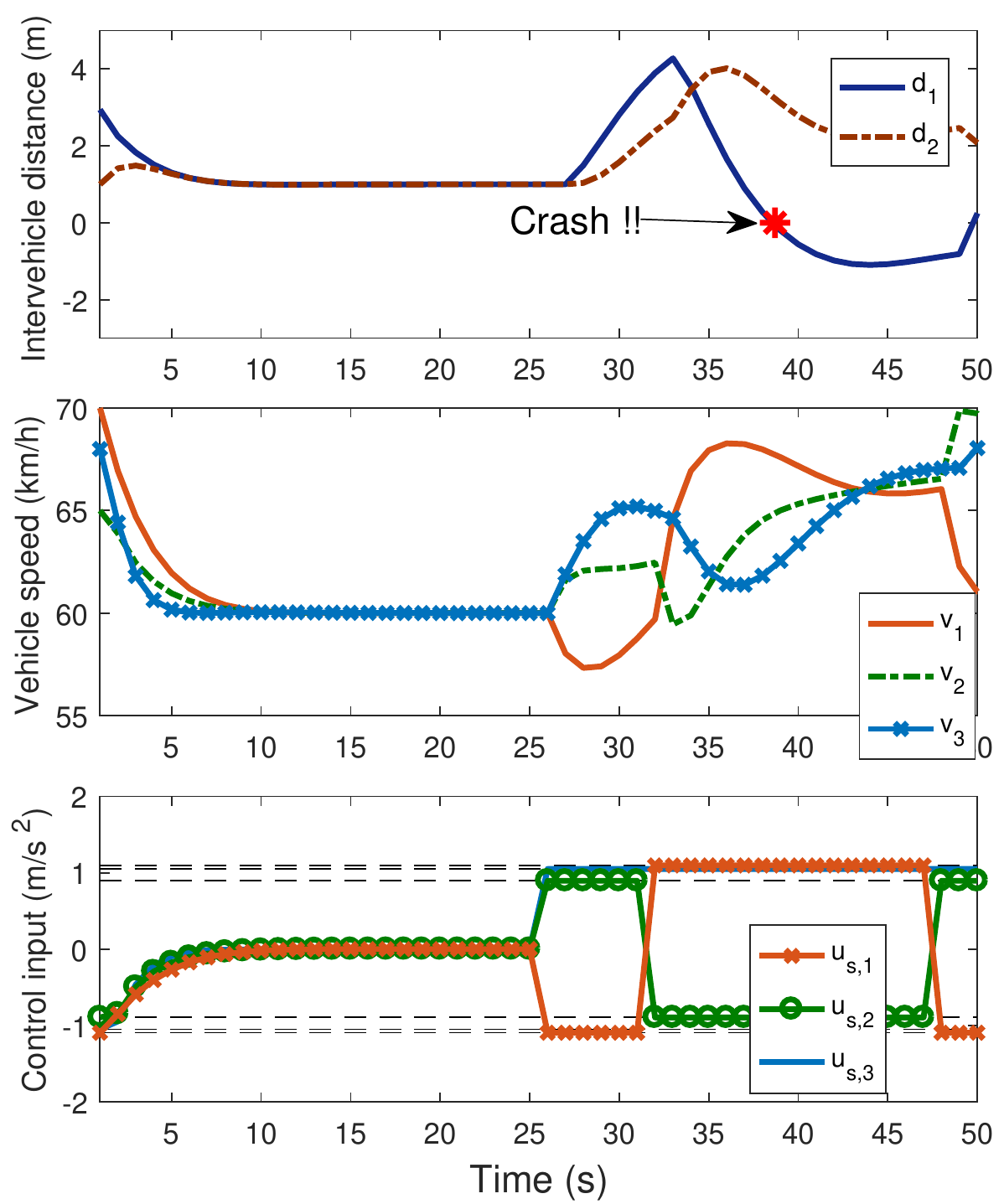}
\caption{Platooning simulations with the CACC  subject to the nominal constraints. An attack is launched after $25\;s$ causing oscillations in the CACC. Due to the sudden changes in the acceleration, the controller cannot maintain a safe distance and vehicles $1$ and $2$ will crash. }\label{fig_attack_original}
\end{figure}

\begin{figure}[t]
\centering
\includegraphics[width=\columnwidth]{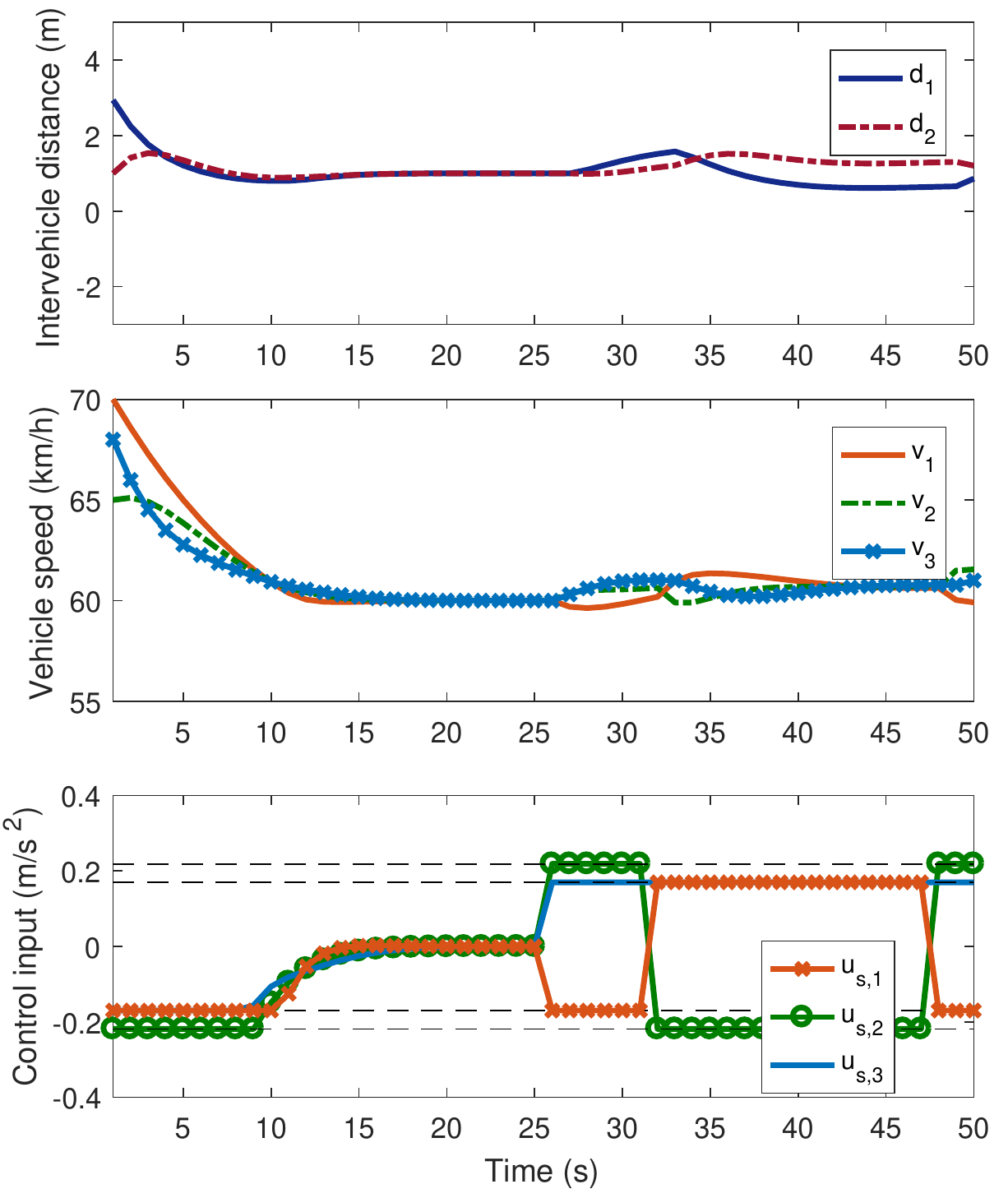}
\caption{Platooning simulations with the CACC  subject to the safe constraints. An attack is launched after $25\;s$ causing oscillations in the CACC. Since the maximum acceleration is bounded with a tighter bound, the vehicles cannot reach unsafe states and so they will never crash.  }\label{fig:secure}
\end{figure}

An adversary gains access to all CACC commands and injects false data that suddenly forces  acceleration/deceleration of the vehicles.    Fig. \ref{fig_attack_original} depicts the distances and velocities when the attack is launched after $25\;$s. Notice that with the original bounds the oscillations provoke a crash between vehicles 1 and 2, i.e., $d_1=0$.  On the other hand, imposing the constraints that we found by applying Theorem \ref{thm:synthesis}, it is possible to prevent the crash as depicted in Fig. \ref{fig:secure}. In fact, any attack in sensors or actuators for any type of secondary control is limited and cannot cause a vehicle crash. 
The cost of these constraints can be observed in the  convergence time. Since the maximum acceleration is small, it takes longer to reach the desired velocity; however, this makes possible to avoid unsafe states.  

\section{CONCLUSION}
In this paper we have taken a new approach to limiting the capabilities of an attacker by imposing artificial limits on the control inputs that drive the system. Whether caused by manipulation of the control inputs themselves, or an indirect effect of sensor or system manipulations, these actuator bounds restrict the states that can be reached. We derive methods based on convex optimization to quantify the reachable states given known actuator bounds and also methods to design new bounds to avoid the reachable set from entering a set of states determined to be unsafe or dangerous. Through the example of a platoon, we show how dangerous states can be determined (e.g., a crash of adjacent vehicles) and avoided. Also, the platoon example demonstrates that security through actuator bounds might come at the cost of conventional performance metrics such as settling time. In future work, we will analyze the  balance between stabilizing the system and securing it with the proposed  bounds.  

\bibliographystyle{IEEEtran}
\bibliography{security}

\end{document}